\documentclass{article}
\usepackage{ijcai21}

\usepackage{times}

\usepackage{soul}
\usepackage{url}
\usepackage[hidelinks]{hyperref}
\usepackage[utf8]{inputenc}
\usepackage[small]{caption}
\usepackage{graphicx}
\usepackage{amsmath}
\usepackage{booktabs}
\title{Reversible Adversarial Attack based on Reversible Image Transformation}

\author{
Zhaoxia Yin$^1$\and
Hua Wang$^1$\and
Li Chen$^1$\And
Jie Wang$^1$\And
Weiming Zhang$^2$\footnote{Contact Author} \\
\affiliations
$^1$Anhui Provincial Key Laboratory of Multimodal Cognitive Computation, Anhui University, Hefei 230601\\
$^2$School of Information Science and Technology, University of Science and Technology of China, Hefei 230026\\
\emails{zhangwm@ustc.edu.cn}
}

\begin{document}

\maketitle

\begin{abstract}
In order to prevent illegal or unauthorized access of image data such as human faces and ensure legitimate users can use authorization-protected data, reversible adversarial attack technique is rise. Reversible adversarial examples (RAE) get both attack capability and reversibility at the same time. However, the existing technique can not meet application requirements because of serious distortion and failure of image recovery when adversarial perturbations get strong. In this paper, we take advantage of Reversible Image Transformation technique to generate RAE and achieve reversible adversarial attack. Experimental results show that proposed RAE generation scheme can ensure imperceptible image distortion and the original image can be reconstructed error-free. What's more, both the attack ability and the image quality are not limited by the perturbation amplitude.
\end{abstract}

\section{Introduction}
In order to make the research significance and technical basis of the proposed work clear, we make introduction the following four aspects. The first is  the research background, leading to the important value of adversarial examples with both attack capability and reversibility. Then, the research status of adversarial attack with  adversarial examples come. After the parallels and differences between information hiding and adversarial examples,  reversible adversarial attacks based on information hiding put forward. Finally, the motivation and contribution of the proposed method is highlighted.

\subsection{Background}
\begin{figure}
\includegraphics[width=\columnwidth]{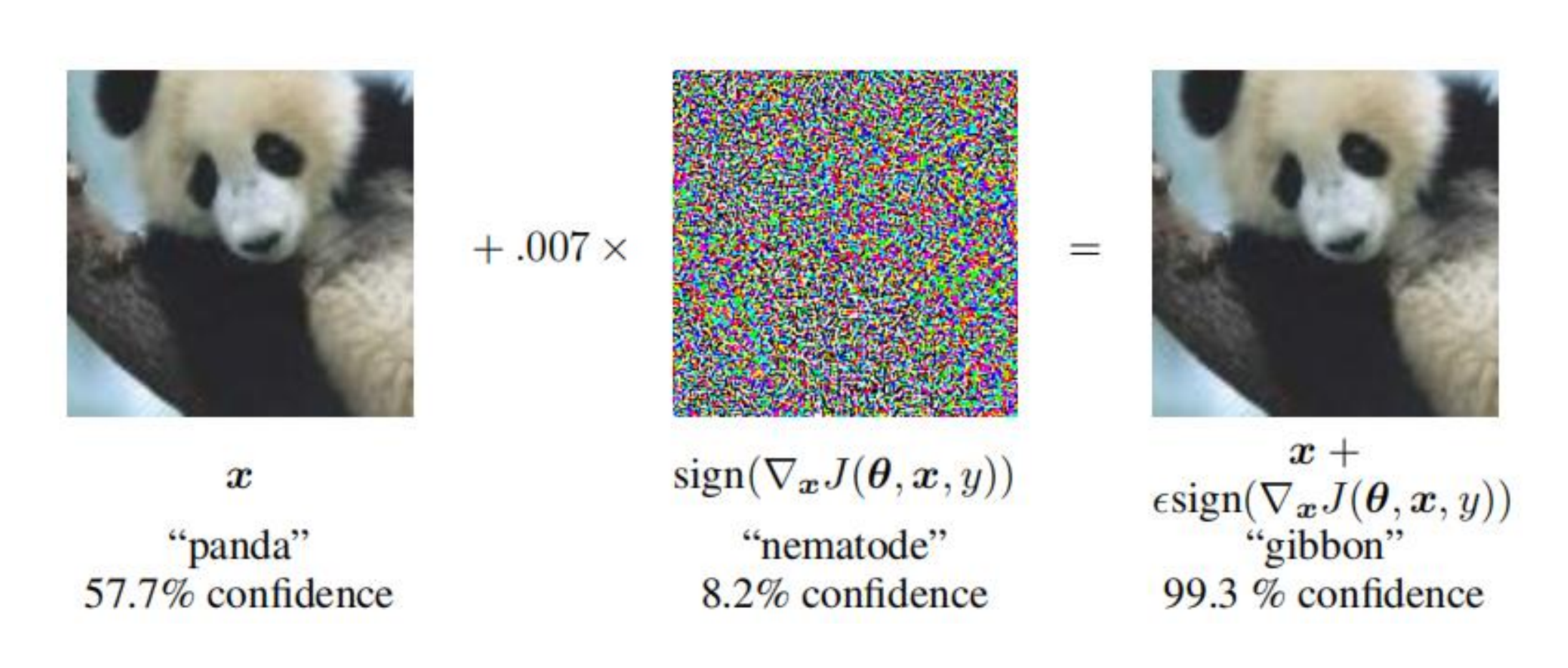}
\caption{The generation process of an adversarial example. } \label{img1}
\end{figure}

Deep learning \cite{LeCun2015Deep} performance is getting more and more outstanding, especially in many tasks such as autonomous driving \cite{aradi2020survey} and face recognition \cite{choi2019ensemble}. As an important technique of AI, it has also been challenged by different kinds of attacks. In 2013, Szegedy et al. \cite{szegedy2013intriguing} first discovered that adding perturbations that are imperceptible to human vision in an image can mislead the neural network model to get wrong results with high confidence. As shown in Fig. \ref{img1}, This kind of images that have been added with specific noise to mislead a deep neural network model are called Adversarial Examples \cite{Goodfellow2014Explaining}, and the added noises are called Adversarial Perturbations.

As a lethal attack technology in the AI security field, if adversarial examples are equipped with both attack capability and reversibility, it will be undoubtedly having important application value, i.e., attacking unauthorized models and harmless to authorized models with lossless recovery capability \cite{hou2019emerging}. 

Reversible adversarial attack aims to add adversarial perturbations into images in a reversible way to generate adversarial examples. On one hand, the generated Reversible Adversarial Examples (RAE) can attack the unauthorized models and prevent illegal or unauthorized access of image data; on the other hand, authorized intelligent system can restore the corresponding original images from RAE completely and avoid interference safely. 
The emergence of RAE equips adversarial examples with new capabilities, which is of great significance to further expand the attack-defense technology and applications of AI. 

However, the research has just started, and the performance are not satisfied. Many problems and questions, such as how to balance and optimize attack capability, reversibility and image visual quality, are still waiting to be solved and answered.

\subsection{Adversarial Attack and Adversarial Examples}
Attacks and defenses of adversarial examples have attracted more and more attention from researchers in the field of machine learning security, and have become a hot research topic in recent years. Here we briefly summarize the current research status of adversarial attack and adversarial examples\cite{2019Adversarial}.

Adversarial attack is to design algorithms to turn normal samples into adversarial examples to fool AI system. According to the different degree of attacker's understanding of the target model information, it can be divided into white box and black box attacks. White-box attack refers to the construction of adversarial examples based on information such as the structural parameters of the target model, Eg. Iterative Fast Gradient Sign Method (IFGSM) \cite{kurakin2016adversarial}. Black-box attack is to construct adversarial examples without any information of the target model and adversarial examples are usually generated by training alternative models, Eg. single pixel attack \cite{2019One}. Further more, taking image classification as an example, non-target attack only needs to make the model result in wrong classification for a given adversarial example and usually the perturbation is relatively small. For example, DeepFool attack \cite{moosavi2016deepfool}.
The other kind of attack can make the model classify a given adversarial example into a specified category rather than any incorrect categories and the representative algorithm is well known as C\&W \cite{carlini2017towards}.
So we can say, by slightly modifying the input digital image signal, adversarial example are generated to show different information to machine or intelligent system. But for human vision, the information and content of the image have not been changed. 

\subsection{Reversible Adversarial Examples}
So we can say, by slightly modifying the input digital image signal, adversarial example are generated to show different information to machine or intelligent system. But for human vision, the information and content of the image have not been changed. Actually, there is another similar technique that also aims to achieve some special goals by slightly modifying the input digital image signal, called Information Hiding, which consists of different research topics such as Watermarking, Steganography and Reversible Data Hiding (RDH) \cite{hou2019emerging}.  

Quiring et al. \cite{schottle2018detecting} analyzed the similarities and differences between Adversarial Example and Watermarking. Both of them modify the target object to cross the decision boundary at the lowest cost. In watermarking, the watermarking detector is regarded as a two-classifier, and the watermarking in the signal could be destroyed by the watermarking attacks, so that the classification result could be changed from image-with-watermarking to image-without-watermarking. In machine learning, this boundary separates different categories, and the attacked signal, i.e. Adversarial Examples, will be misjudged by the model. Schöttle et al. \cite{quiring2018forgotten} analyzed the similarities and differences between steganography and adversarial examples. Steganography attempts to modify individual pixel values to embed secret information, so that it is difficult for steganography analysts to detect the hidden information. Schöttle et al. believe that the detection of adversarial examples belongs to the category of steganalysis, and develops a heuristic linear predictive adversarial detection method based on steganalysis technology. Zhang et al. \cite{zhang2021universal} compared deep steganography and universal adversarial perturbation, and found that the success of both is attributed to the deep neural network’s exceptional sensitivity to high frequency content.

When we know these interesting cross-cutting studies of Adversarial Example and Information Hiding, we would inevitably wonder, what would we get by combining Adversarial Example with another Information Hiding technique, i.e. Reversible Data Hiding.

\begin{figure}
\includegraphics[width=\columnwidth]{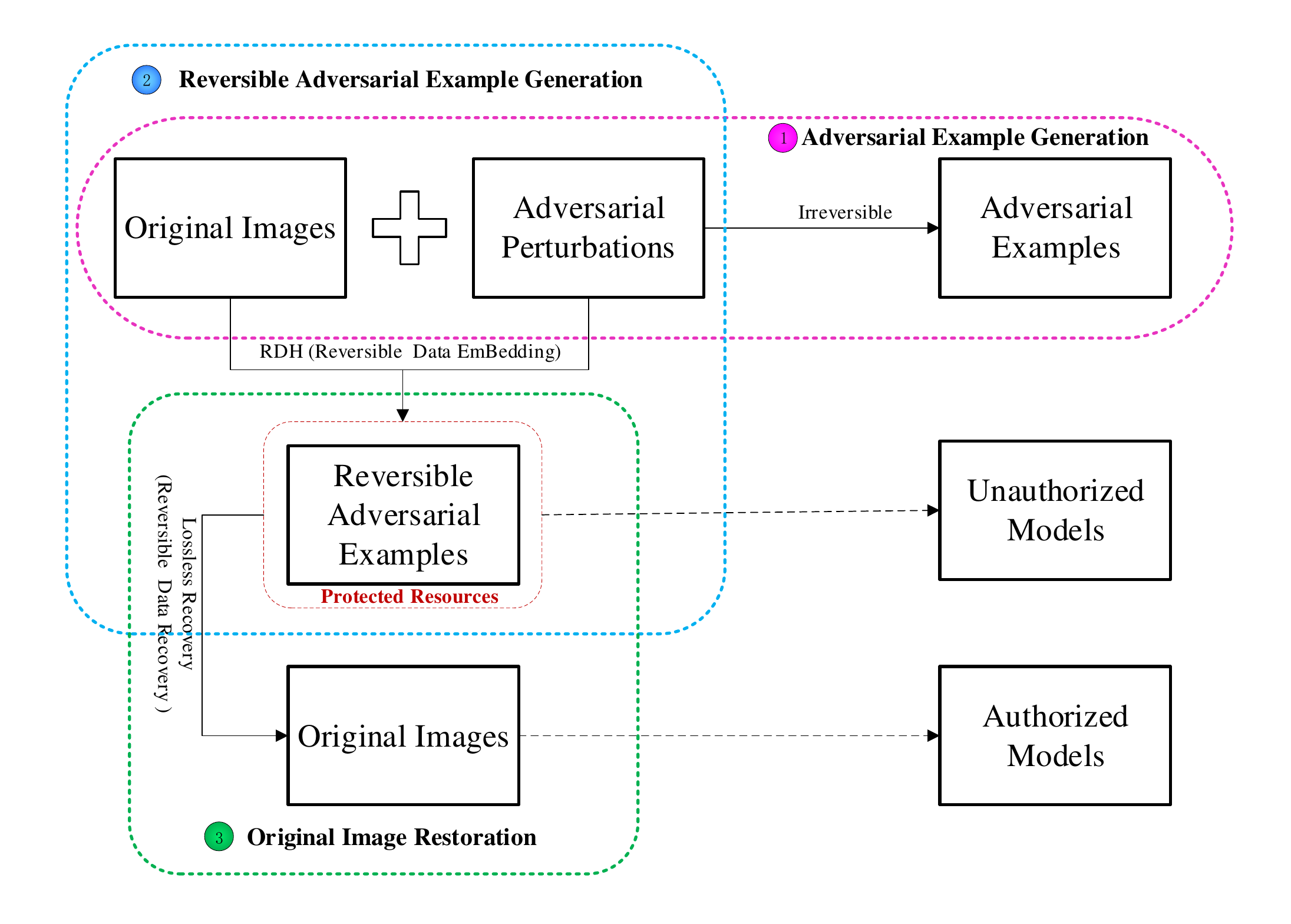}
\caption{The overall framework of RDE-based RAE method \protect\cite{liu2018reversible}. } \label{img2}
\end{figure}

Liu et al. achieved the first Reversible Adversarial attack by combining Reversible Data Hiding with Adversarial Examples and proposed the concept of Reversible adversarial examples (RAE) \cite{liu2018reversible}. Since RAE get both attack capability and reversibility at the same time, illegal or unauthorized access of image data can be prevented and legitimate using can be guaranteed by original image recovery. As shown in Fig. \ref{img2}, Reversible Data Embedding (RDE) technique \cite{zhang2013recursive} is adopted to embed the adversarial perturbations into its adversarial image to get the reversible adversarial example image, from which, the original image can be restored error-free. The framework consists of three steps: (1) adversarial examples generation; (2) reversible adversarial examples generation by reversible data embedding; (3) original images recovery. 
This is really a great creative work even though the performance is far from satisfied. Let's call it RDE-based RAE method and discuss the details in the coming section.

\subsection{Motivation and Contribution}
As mentioned above, to obtain RAE, Liu et al. adopted Reversible Data Embedding technique to embed the adversarial perturbations into its adversarial image, then the original image can be restored without distortion.

However, no matter which kind of RDE algorithm is adopted, the embedding capacity is always limited.
That means, the maximum amount of the embedding data that can be carried by the adversarial image is also limited. 
Therefore, when adversarial perturbations are strengthened, the amount of data that needs to be embedded increases, that would result in the following three problems: (1) The generated adversarial perturbations cannot be embedded completely and then the original image cannot be restored completely , which leads to the failure of reversibility; (2) Since too much data has to be embedded, the reversible adversarial image is severely distorted, which leads to unsatisfied image quality; (3) Due to increased distortion of RAE, the attack ability decreases accordingly.

To solve these problems, here we propose to replace the idea of Reversible Data Embedding with Reversible Image Transformation (RIT) technique. In order to verify the effectiveness of the strategy, we chose one RIT method \cite{hou2018reversible} as an example to construct RAE and make performance comparisons with the method from \cite{liu2018reversible}. Experiments show that the proposed scheme can completely solve the problems that analyzed above. Furthermore, in the proposed method, realization of reversibility does not depend on embedding the signal difference between original images and adversarial examples, i.e., it is not limited to the strength of adversarial perturbations. As well-known, the greater the adversarial perturbation, the stronger the attack ability. Therefore, the proposed method can achieve better RAE performance in terms of reversibility, image quality and attack capability. We name it RIT-based RAE method and describe it step by step in Section 2. Details of experiments and results are given in Section 3, following with Conclusion in Section 4.

\section{The Proposed Method}
\begin{figure} %[t] 
\centering 
\includegraphics[width=\columnwidth]{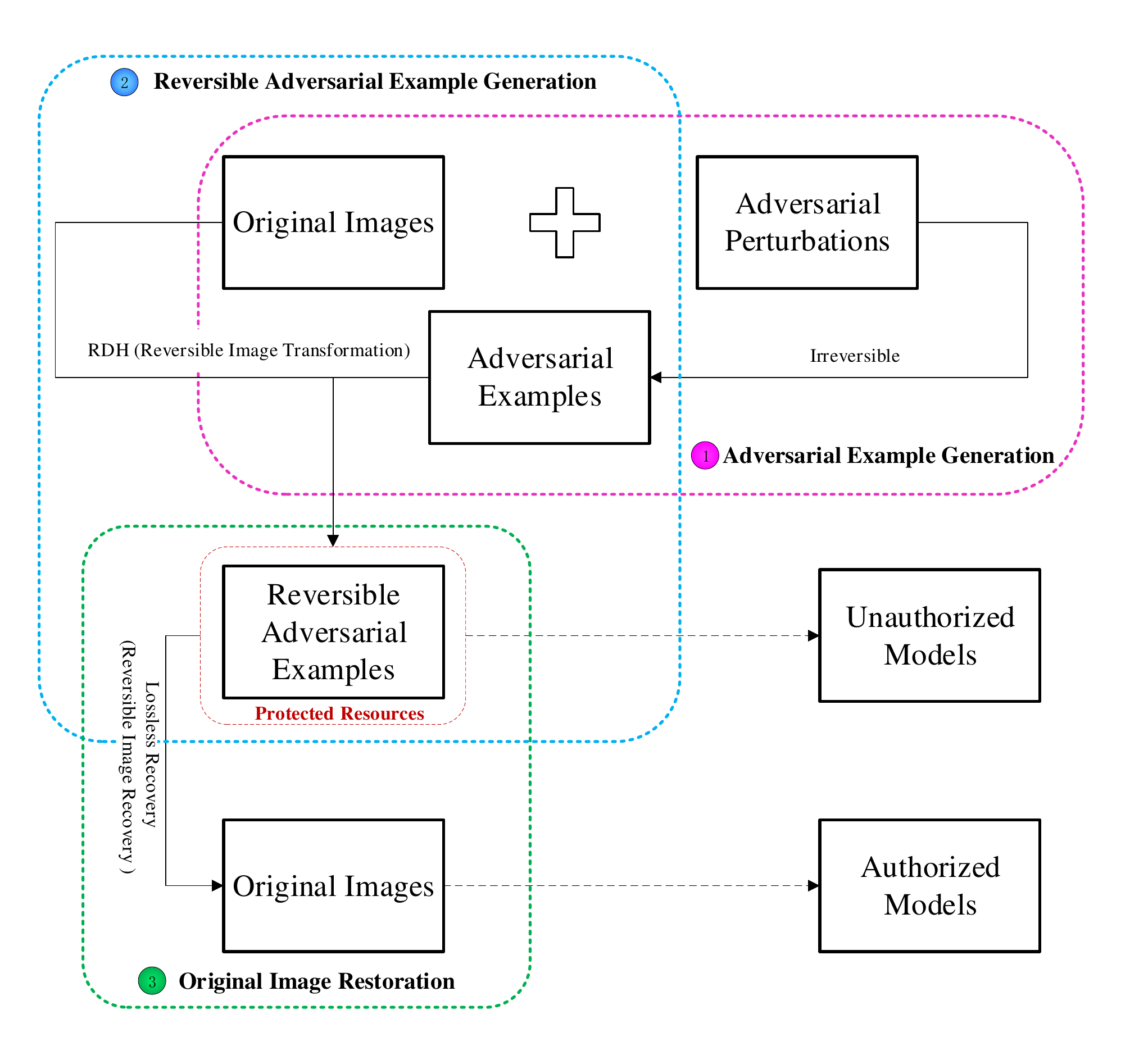}
\caption{The overall framework of the proposed RIT-based RAE method.} \label{img3}
\end{figure}

In order to achieve reversible adversarial attack, we propose a more effective method to generate reversible adversarial examples. As shown in Fig. \ref{img3}, we replace reversible data hiding with RIT strategy to obtain RAE. The original image restoration process is the inverse process of RIT, i.e., reversible image recovery. In this section, We describe the implementation of our method as three steps: (1) Adversarial examples generation; (2) Reversible adversarial examples generation; (3) Original image restoration.

\subsection{Adversarial Examples Generation}
Firstly, we need to generate  adversarial examples  for step (2). Since adversarial attacks are mainly divided into white box and black box. White box attack algorithms have better performance, and black box attacks usually rely on white box attacks indirectly, so this paper generates adversarial examples based on white-box settings. Next, we introduce several state-of-the-art white box attack algorithms.

\begin{itemize}

\item {\textbf{IFGSM}} \cite{kurakin2016adversarial}  proposed as an iterative version of FGSM  \cite{Goodfellow2014Explaining}. It is a quick way to generate adversarial examples, applies FGSM multiple times with small perturbation instead of adding a large perturbation. 

\item{\textbf{DeepFool}}\cite{moosavi2016deepfool} is a untargeted attack algorithm that generates adversarial examples  by exploring the nearest decision boundary, the image is slightly modified in each iteration to reach the boundary, and the algorithm will not stop until the modified image changes the classification result. 

\item{\textbf{C\&W}} \cite{carlini2017towards} is an optimization-based attack that makes perturbation undetectable by limiting the $L_0$, $L_2$, $L\infty$ norms. 

\end{itemize}

\subsection{Reversible Adversarial Examples Generation}

Secondly, we take Reversible Image Transformation (RIT) algorithm to generate protected resources with restricted access capabilities, i.e., reversible adversarial examples. Specifically, we take the adversarial example as the target image, and use RIT to disguise original image as the adversarial example to directly get reversible adversarial example. Next, we will introduce the RIT algorithm in detail. In fact, RIT algorithm is also a kind of reversible data hiding technique to achieve image content protection. It can reversibly transform an original image into an arbitrarily-chosen target image with the same size to get a camouflage image, which looks almost indistinguishable from the target image. When the difference between the two images is smaller, the amount of auxiliary information for restoring original image is greatly reduced, that makes it perfect for RAE since the difference between an original image and its Adversarial Example is usually very small.

\subsubsection{Algorithm Implementation}
In order to facilitate the understanding of the implementation of the RIT algorithm, the following takes grayscale image (one channel) as an example to illustrate the specific implementation process of the algorithm \cite{zhang2016reversible}. For color images, the R, G, and B color channels are transformed in the same way. RIT achieves the reversible transformation on two pictures, and there are two stages of transformation and restoration,  in the transformation stage, the original image undergoes a series of pixel value transformations to generate a camouflage image \cite{zhang2016reversible}. In the recovery stage, the hidden image transformation information needs to be extracted from the camouflage image, and is used for reversible restoration. Since the restoration is the reverse process of the transformation, we only need to introduce the transformation process. The transformation process is divided into three steps: (1) Block Paring (2) Block Transformation (3) Accessorial Information Embeding.

\begin{itemize}
\item {\textbf{Block Paring}}
The original image and the target image are divided into blocks in the same way firstly. Then, calculate Mean and Standard Deviation of the pixel values of each block of the original image and the target image. Next, to restore the original image from the camouflage image, the receiver must know the Class Index Table of the original image. By matching the blocks in the original image with the blocks in the target image with similar Standard Deviations into a pair, the original image and the target image can be obtained separately Class Index Table.

\item {\textbf{Block Transformation}}
Firstly, according to the block matching method, each pair of blocks has a close Standard Deviation value. Do not change the Standard Deviation of the original image, just change the mean value of the original image through the average shift. Then, in order to keep the similarity between the transformed image and the target image as much as possible, further rotate the transformed block to one of four directions: 0$^{\circ}$, 90$^{\circ}$, 180$^{\circ}$, 270$^{\circ}$, and choose the best direction to minimize Root Mean Square Error between the rotating block and the target block.

\item {\textbf{Accessorial Information Embeding}}
In order to obtain the final camouflage image, it is necessary to embed auxiliary information into the transformed image, including: compressed Class Index Table and the average shift and rotation direction of each block of the original image. Choose a suitable RDH algorithm embeds these auxiliary information into the transformed image to get the final camouflage image.

\end{itemize}

\subsection{Original Image Restoration}
Finally, the original image needs to be restored when an unauthorized model accesses it, the restoration process of RIT can be directly used to realize the reverse transformation of the reversible adversarial example to the original image. Since our reversible adversarial examples are based on RIT technology, the process of restoring the reversible adversarial examples to the original image is the restoration process of RIT, while the restoration process is the inverse process of the RIT transformation process. Therefore, in the case of only reversible adversarial examples, we can extract the hidden transformation information, and take the information to reverse the RIT transformation process to non-destructively restore the original images.

\section{Evaluation and Analysis}
To verify the effectiveness and superiority of the proposed method, here we introduce the experiment design, results and comparisons, following with discussion and analysis. 

\subsection{Experimental Setup}
\begin{itemize}
\item{\textbf{Dataset:}}
Since it is meaningless to attack images that have been mis-classified by the model, we randomly choose 5000 images from ImageNet (ILSVRC 2012 verification set) that can be correctly classified by the model for experiments.
\item{\textbf{Network:}}
The pretrained Inception\_v3 in torchvision.models that is evaluated by Top-1 accuracy.
\item{\textbf{Attack Methods:}}
IFGSM, C\&W, DeepFool. To ensure the visual quality, we set the learning rate of C\&W\_L2 to 0.005, the perturbation amplitude $\epsilon$ of IFGSM no more than 8/225.
\end{itemize}

\subsection{Performance Evaluation}
In order to evaluate the performance of the proposed method, we measure attack success rates as well as image quality of our reversible adversarial examples, and compare our RIT-based RAE method with RDE-based RAE method proposed by of Liu et al. \cite{liu2018reversible}.

In order to detect the attack ability of the generated reversible adversarial examples, firstly, three white-box attack algorithms are taken to attack the selected original images to get adversarial examples. Then, we take reversible image transformation to transform original images into target adversarial images to generate reversible adversarial examples. Finally, we utilize the generated reversible adversarial images to attack the model to get attack success rates. As shown in Table \ref{tab3}, the second line shows the attack success rates of the generated adversarial examples (which are non-reversible). The third and fourth lines are the attack success rates of Liu et al.'s and our reversible adversarial examples under different settings, respectively. On IFGSM, when $\epsilon$ is 4/225, 8/225, the attack success rates of our RAEs are: 70.80\%, 94.55\% respectively. In the same case, the attack success rates of Liu et al.'s RAEs are only 35.22\%, 81.00\%, respectively. On C\&W\_L2, when confidence $\kappa$ is 50, 100, the attack success rates of our RAEs are: 81.02\% and 94.84\%, while that Liu et al.'s method are just 52.73\%, 55.01\%, respectively. From the results presented in Table \ref{tab3}, we observe that the attack ability of the RAEs obtained by our method is superior to that of Liu et al's method. But on DeepFool, because the adversarial perturbation generated by this attack closes to the theoretical minimum, its robustness is also relatively poor. Therefore, the amount of information embedded in the adversarial examples generated by the DeepFool exceeds a certain amount, which will seriously weaken the attack performance of the adversarial examples. In this kind of attack algorithm with minimal disturbance and low robustness, the amount of auxiliary information embedded in RIT-based RAEs is greater than the amount of perturbation signal embedded in RDE-based RAEs of Liu et al., so the success rates of our RAEs are lower.

Further more, we found that, when adversarial perturbations get stronger, the amount of data that needs to be embedded increases, which leads to the failure of reversibility for RDE-based RAEs. Take Giant Panda image from Fig.1 as an example, on C\&W, when confidence $\kappa$ is 100, the amount of data that needs to be embedded by RDE-based RAE is 316311 bits, that's far from the corresponding highest embedding capacity 114986 bits. At the same time, to achieve reversible attack by using the proposed RIT-based RAE method, the amount of additional data that needs to be embedded is only 105966 bits.

\begin{table*}
\centering 
\caption{The attack success rates of original adversarial examples, RDE-based RAEs and RIT-based RAEs. }\label{tab3}
\begin{tabular}{lrrrrr}
\hline
&  IFGSM &  IFGSM  & C\&W\_L2 & C\&W\_L2 & DeepFool \\
&  ($\epsilon$=4/225) &  ($\epsilon$=8/225) & ($\kappa$=50) & ($\kappa$=100) & \\
\hline
Generated AEs & 73.84\% & 95.34\%  & 99.98\% & 100\% & 98.35\% \\
\cite{liu2018reversible} & 35.22\% & 81.00\% & 52.73\% & 55.01\% & 84.19\% \\
\textbf{Our RAEs} & 70.80\% & 94.55\% & 81.02\% & 94.84\% & 54.68\% \\
\hline
\end{tabular}
\end{table*}

Then, to quantitatively evaluate the image quality of RAEs, we measure three sets of PSNR: RAEs and original images,RAEs and adversarial examples as well as original images and adversarial examples. The general benchmark for PSNR value is 30dB, and the image distortion below 30dB can be perceived by human vision. In order to make a fair comparison with the method of Liu et al.\cite{liu2018reversible}, we keep the original image and the adversarial example consistent in the experiment, and the corresponding values of PSNR are shown in the last column of Table \ref{tab4}. By comparing RAEs on IFGSM and C\&W with the original images, we found that the PSNR values of our method are higher, that means the generated RAEs are less distorted than that of Liu et al. The comparison between the RAEs and the original adversarial examples shows that our PSNR values are basically greater than 30dB, indicating our RAEs are closer to the original adversarial examples. This result is also consistent with the data in Table 1, that means the specific structure of adversarial perturbation is better preserved in our method, so that the final RAEs have almost the same attack effect as the original adversarial example on IFGSM and C\&W. Similar to the experimental data in Table 1 again, for attack algorithms like DeepFool, the perturbation embedding amount in Liu et al.'s method is smaller than the auxiliary information embedding amount in our work, so the PSNR values of our RAEs are smaller.

\begin{figure} 
\centering 
\includegraphics[width=\columnwidth]{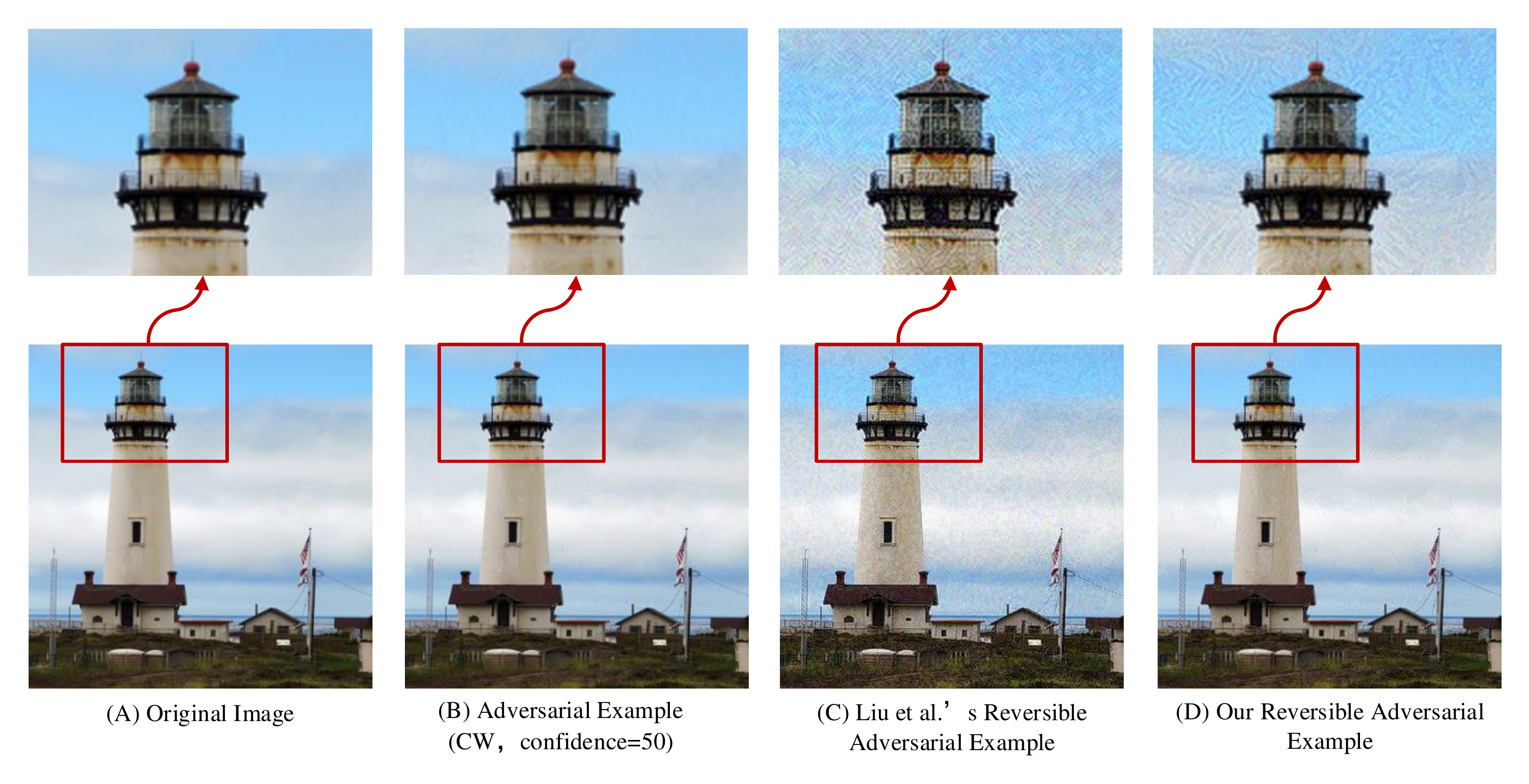}
\caption{Sample figures of reversible adversarial examples generated by different methods.} \label{img5}
\end{figure}

 In addition, Fig.\ref{img5} shows the sample images of RAEs generated by Liu et al. and our method, respectively. After partial magnification, we can see that the image distortion of RDE-based RAEs significantly exceeds that of RIT-based RAEs. Since the amount of auxiliary information embedded in RIT-based RAEs is relatively stable, while the amount of perturbation embedded in RDE-based RAEs is related to the perturbation signal. The greater the perturbation, the more the amount of information embedded, and the more the image distorted.

\begin{table*} 
\centering 
\caption{Comparison results of image quality with PSNR(dB). }\label{tab4}
\begin{tabular}{lrrrr}
\hline
Attacks & Methods & RAEs/OIs & RAEs/AEs & OIs/AEs \\
\hline
IFGSM($\epsilon$=4/225) & \cite{liu2018reversible} & 22.64 & 23.26 & 37.69 \\
& \textbf{Proposed method} & 30.81 & 33.15 &  \\
\hline
IFGSM($\epsilon$=8/225) & \cite{liu2018reversible} & 21.93 & 23.55 & 32.31 \\
& \textbf{Proposed method} & 27.59 & 32.13 &  \\
\hline
C\&W\_L2($\kappa$=50) & \cite{liu2018reversible} & 26.15 & 26.40 & 44.66  \\
& \textbf{Proposed method} & 33.64 & 35.09 &  \\
\hline
C\&W\_L2($\kappa$=100) & \cite{liu2018reversible} & 22.57 & 23.07 & 38.83 \\
& \textbf{Proposed method} & 32.13 & 34.87 &  \\
\hline
DeepFool &  \cite{liu2018reversible} & 40.24 & 40.85 & 51.04 \\
& \textbf{Proposed method} & 34.44 & 35.48 &  \\
\hline
\end{tabular}
\end{table*}

%RAEs

\subsection{Discussion and Analysis}
Both RDE-based RAE and RIT-based RAE use RDH technology to achieve adversarial reversible attacks. In RDE-based RAE framework, Liu et al. take reversible data embedding algorithm to hide the perturbation difference in the adversarial example to get a reversible adversarial image. Constrained by the RDH payload, to achieve reversibility, the perturbation signal can only be controlled within the range of the payload. A slight increase in the perturbation amplitude will cause serious visual distortion of the reversible adversarial example, severely weakened attack ability, and even unable to fully embed the perturbation signal so that the original image cannot be restored reversibly. In the proposed RIT-based RAE framework, since reversible image transformation is unnecessary to consider the size of the adversarial perturbation, the problem of difficulty in embedding adversarial perturbation is solved, and it further improves the visual quality of the reversible adversarial example to promote the overall attack success rates. In a sense, the attack effect of our reversible adversarial examples is affected to a certain extent by the amount of auxiliary information needed to restore the original image, and the amount of auxiliary information is usually relatively stable. Generally speaking, we can reduce the impact of auxiliary information embedding by enhancing the adversarial perturbation. That is to say, when generating an adversarial image, the robustness of the adversarial example is improved by increasing the perturbation amplitude, and finally the attack success rate of the generated reversible adversarial example is improved. However, when faced with an attack algorithm similar to DeepFool with less perturbation and low robustness, since RIT auxiliary information embedding has a greater impact on its performance than perturbation signal embedding, the attack success rate of our reversible adversarial examples is lower than Liu et al. While the proposed scheme is a special application of RIT, original image and its target adversarial image have a high degree of similarity. Our future work is to improve reversible image transformation algorithm based on the similarity between original image and adversarial example so that the attack success rate of reversible adversarial examples is further improved. 

\section{Conclusion}
To solve the problems of RAE technique and improve the performance in terms of reversibility, image quality and attack ability, we take advantage of reversible image transformation to construct reversible adversarial examples, which aims to achieve reversible attack. In this work, we regard a generated adversarial example as the target image and its original image can be disguised as its adversarial example to get RAE. Then the original image can be recovered from its reversible adversarial example without distortion. Experimental results illustrate that our method overcomes the problems of perturbation information embedding. Moreover, it's even achieved that the larger adversarial perturbation, the better RAE can be generated. RAE can prevent illegal or unauthorized access of image data such as human faces and ensure legitimate users can use authorization-protected data. Today, when deep learning and other artificial intelligence technologies are widely used, this technology is of great significance. In future work, it is worth trying to further combine more reversible information hiding technologies to study RAE solutions that meet actual needs.

\bibliographystyle{named}
\bibliography{ijcai21}

\end{document}